\pdfoutput=1

\documentclass[aps,prx,reprint,showpacs,superscriptaddress,floatfix]{revtex4-1}
\usepackage{graphicx}
\usepackage{graphics}
\usepackage{amsmath}
\usepackage{amssymb}
\usepackage{amsfonts}
\usepackage{dcolumn}
\usepackage{dsfont}
\usepackage{latexsym}
\usepackage{rotating}
\usepackage{color}
\usepackage{latexsym}
\usepackage{bbm}
\usepackage{subfigure}
\usepackage{float}
\usepackage{epsfig}
\usepackage{epsf}
\usepackage{psfrag}
\usepackage{bm}
\usepackage{amsthm}
\usepackage{eucal}
\usepackage{mathrsfs}
\usepackage{url}
\usepackage{braket}
\usepackage{array}
\usepackage[utf8]{inputenc}
\usepackage{microtype}
\usepackage{multirow}
\usepackage{comment}
\usepackage{hyperref}
\hypersetup{
colorlinks=true,final=true,
        linkcolor=blue,
        citecolor=blue,
        filecolor=blue,
        urlcolor=blue,
}

\begin{document}
\title{Pressure effects on the electronic structure and magnetic properties of infinite-layer nickelates}
\author{Shekhar Sharma}
\email{sshar246@asu.edu}
\affiliation{Department of Physics, Arizona State University, Tempe, AZ 85287, USA}
\author{Myung-Chul Jung}
\affiliation{Department of Physics Education, Chosun University, 30 Chosundae3gil, Dong-gu, Gwangju, South Korea, 61452}
\author{Harrison LaBollita}
\affiliation{Department of Physics, Arizona State University, Tempe, AZ 85287, USA}
\author{Antia S. Botana}
\affiliation{Department of Physics, Arizona State University, Tempe, AZ 85287, USA}

\date{\today}
\begin{abstract}

Motivated by the discovery of superconductivity in infinite-layer nickelates RNiO$_2$ (R= rare-earth), and the subsequent enhancement of their T$_c$ with pressure, we investigate the evolution of the electronic structure and magnetic properties of this family of materials via first-principles calculations employing hydrostatic and chemical pressure as tuning knobs.  
Overall, our analysis shows that pressure tends to increase the R-$5d$ self-doping effect, as well as the Ni-$d
_{x^{2}-y^{2}}$ bandwidth, the $e_g$ energy splitting, the charge transfer energy, and the superexchange ($J$).  Using the energy scale of $J$ as a predictor of superconducting tendencies, we anticipate that pressure can indeed be a feasible means to further increase the T$_c$ in this family of materials.
\end{abstract}

\maketitle

\section{\label{sec:intro}Introduction}
The discovery of high-T$_c$ superconductivity (HTS) in the cuprates~\cite{Bednorz1986} triggered attempts to find analog superconducting materials aimed at identifying the requisite ingredients for HTS \cite{Keimer2015}. In this context, nickelates have been intensively investigated given the proximity of Ni to Cu in the periodic table~\cite{Ni1+isnotCu2+, anisimov1999}. Nickelate superconductivity was first realized in 2019 in hole-doped infinite-layer Nd$_{1-x}$Sr$_{x}$NiO$_{2}$ thin films with T$_c\sim15$ K~\cite{2019infinite_layer_superconductivity}. This initial discovery immediately attracted extensive theoretical and experimental efforts \cite{Osada2020_nanolett, hepting2020, Li2020dome, BiXiaWang2020, Osada2021_advmat, wu2020_PRB, Hu2019, jiang2019, PRL_Jiang_Ni_spin_state_2020, nomura2019, Choi2020_PRB, choi2020_PRR, Ryee2021_prl, Ryee2020, Gu2020, Karp2020_112, Karp2021, Leonov2020, lechermann2020late, lechermann2020multi, Sakakibara2020, jiang2019, wang2020, petocchi2020, werner2020, Gu2020single,Krishna2020effects, kitatani2020,Kang2021opt, lee2020aspects, Li2020, goodge2020} that gave rise to the subsequent discovery of superconductivity in hole-doped infinite-layer nickelates with other rare-earth cations~\cite{Osada2020_prm,Zeng2021superconductivity,Osada2021_advmat}, in a quintuple-layer nickelate (Nd$_{6}$Ni$_{5}$O$_{12}$) without chemical doping~\cite{pan2021super}, and most recently in the parent Ruddlesden-Popper bilayer \cite{bilayer1, bilayer2, bilayer3} and trilayer \cite{trilayer1, trilayer2, trilayer3} nickelates upon applying hydrostatic pressure.

Focusing on the infinite-layer nickelates, these materials exhibit striking similarities to the cuprates. 
Structurally, both materials host two-dimensional square planes of transition-metal and O atoms \cite{crespin1983,hayward1999}. In terms of filling, both are in proximity to a 3$d^9$ electronic configuration with superconductivity arising near  20$\%$ hole doping.
The $e_g$ splitting -- correlated with T$_c$ in the cuprates, with a larger value giving
rise to a higher T$_c$ due to reduced mixing of these
orbitals \cite{sakakibara_splitting, Sakakibara2014} -- is also similar in infinite-layer nickelates~\cite{botana2020}.  Theoretical and experimental work suggests an unconventional ($d$-wave) pairing mechanism for superconductivity in these nickelates, akin to the cuprates \cite{wu2020_PRB, kitatani2020, Cheng2024, harvey2022evidence}. Further, a recent analysis of pressure effects on hole-doped PrNiO$_2$ thin films \cite{wangnn_pressure_enhanced} has revealed a substantial increase in T$_c$ (from 17 K to 31 K) under 12 GPa, following the pressure-enhanced T$_c$ trends of the cuprates \cite{cuprate_pressure}. In spite of these similarities, there are, however, some important differences between infinite-layer nickelates and cuprates that have been intensively studied ~\cite{Ni1+isnotCu2+,botana2020}. The cuprates are antiferromagnetic charge-transfer insulators close to the $3d^9$ limit that portray a single band of $d_{x^2-y^2}$ character near the Fermi level, and are known to exhibit a large degree of O-$p$ and Cu-$d$ hybridization \cite{Keimer2015}. For infinite-layer nickelates, the lower degree of hybridization between the Ni-$d$ and O-$p$ states, as well as the presence of additional `spectator' or `self-doping' bands of rare-earth (R)-$d$ character at the Fermi level are distinguishing factors \cite{botanareview}. These rare-earth electron pockets that self-dope the Ni-$d_{x^{2}-y^{2}}$ band preempt the possibility of an antiferromagnetic insulating state in the parent phase, even though the presence of strong antiferromagnetic correlations has
recently been reported via resonant inelastic x-ray scattering
(RIXS) experiments \cite{Lu2021}.

In light of the experimental enhancement of T$_c$ in infinite-layer nickelates with pressure (together with the recent explosion of work on superconductivity in pressurized Ruddlesden-Popper nickelates), here,  we study the effects of hydrostatic pressure on the electronic structure and magnetic response of infinite-layer nickelates. Using first-principles calculations, we report the systematic evolution of the dominant similarities and differences between infinite-layer (RNiO$_2$) nickelates and cuprates with pressure. Further, we correlate the changes arising from hydrostatic pressure with changes in chemical pressure, by exploring different rare-earth ions. We find that both hydrostatic and chemical pressure can be used as a ``knob'' to tune the electronic and magnetic response in infinite-layer nickelates to ultimately enhance their T$_c$.

\begin{figure}
    \centering
    \includegraphics[width=\columnwidth]{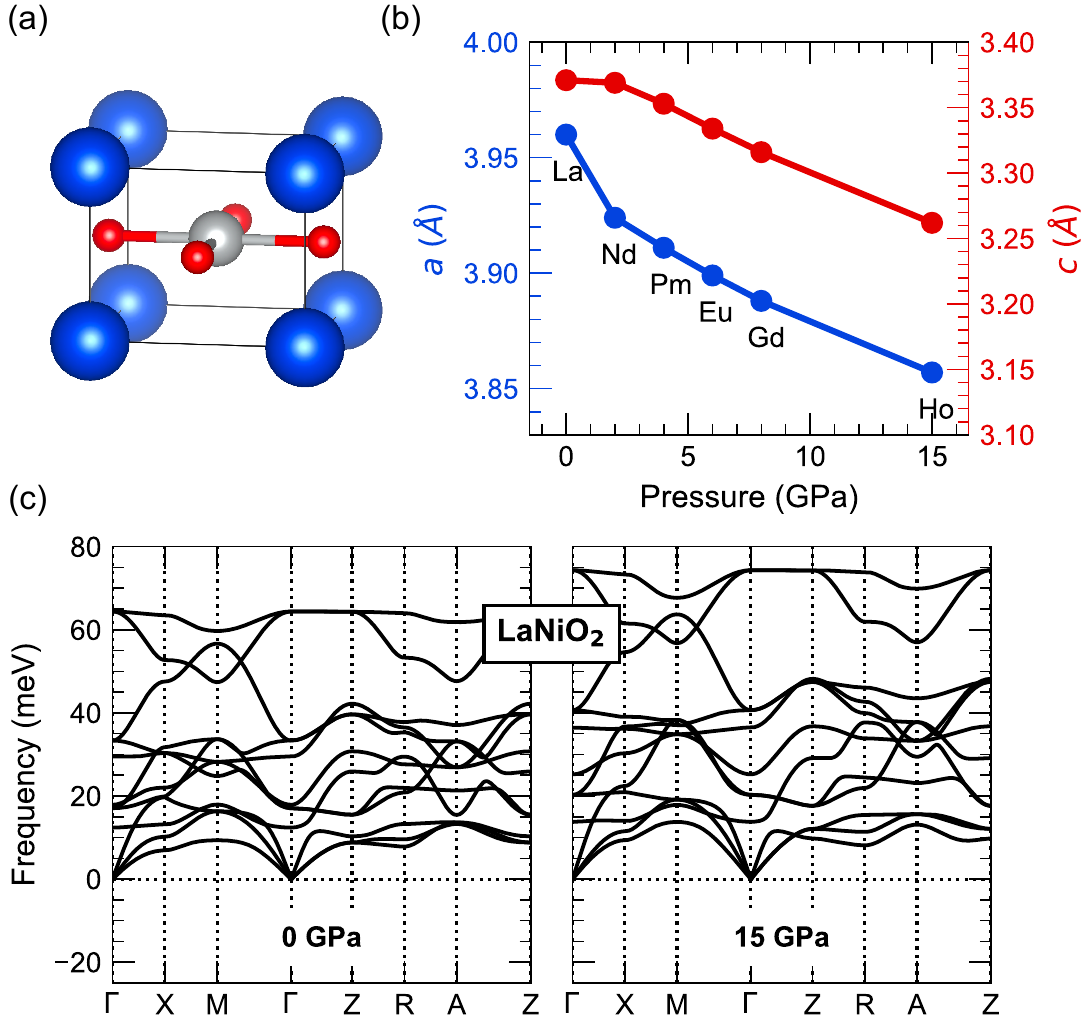}
    \caption{Crystal structure and lattice stability of pressurized infinite-layer RNiO$_{2}$. (a) Crystal structure of RNiO$_{2}$ nickelates ($P4/mmm$) where blue, grey, and red spheres denote R, Ni, and O atoms, respectively. (b) DFT-optimized lattice constants for pressurized LaNiO$_{2}$ with corresponding rare-earth cations matched to the in-plane lattice constant of LaNiO$_{2}$ at a certain pressure. (c) Phonon dispersions for LaNiO$_{2}$ at ambient pressure (left) and 15 GPa (right).}
    \label{fig:lattice}
\end{figure}

\begin{figure*}
    \centering
    \includegraphics[width=2.0\columnwidth]{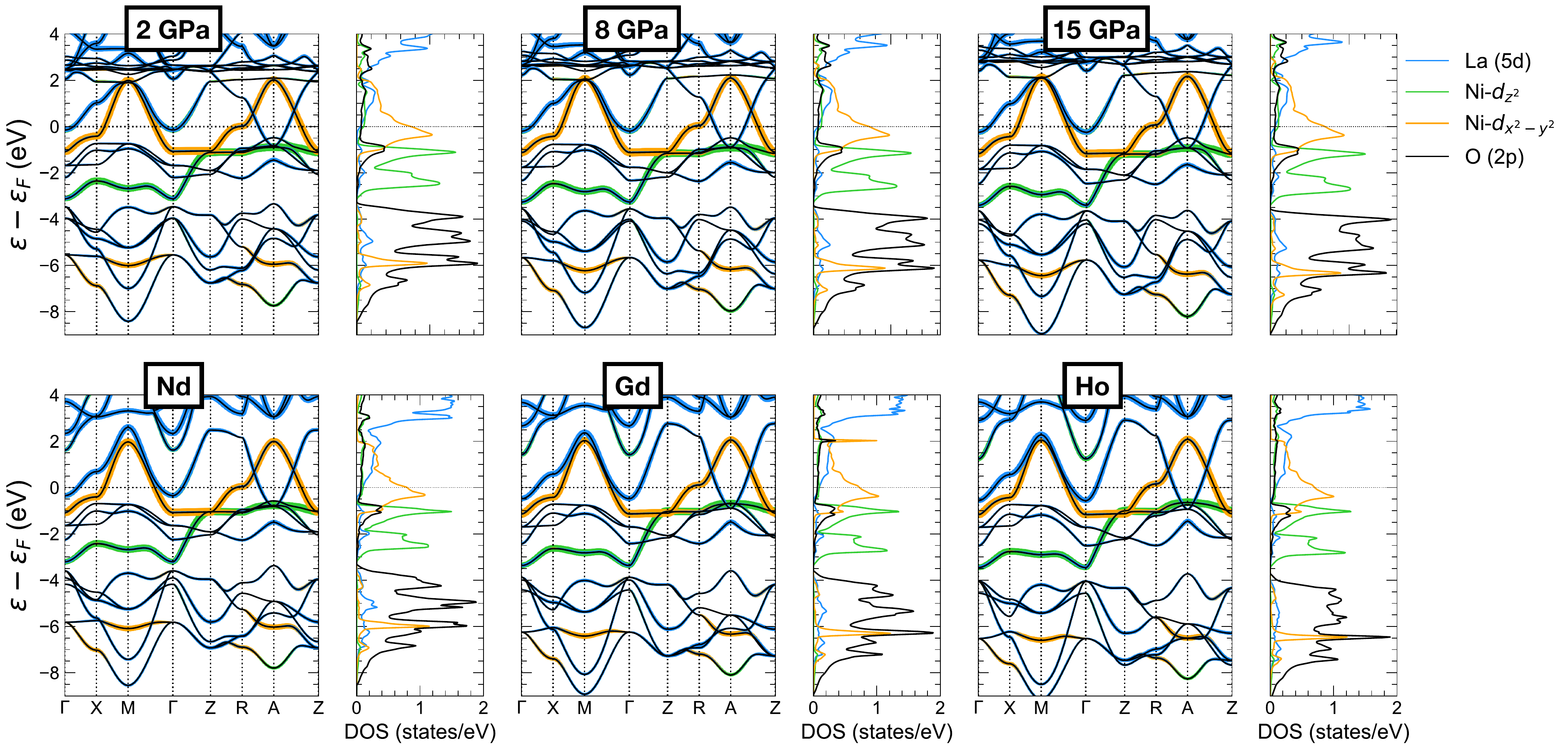}
    \caption{Comparison between the non-magnetic electronic structure of RNiO$_2$ with hydrostatic (top) and chemical (bottom) pressure. Top panels: Band structure along high-symmetry lines and atom-, orbital-resolved density of states (DOS) for LaNiO$_2$ at 2 GPa, 8 GPa, and 15 GPa (from left to right). The orbital character of the bands is denoted for Ni-$d_{x^{2}-y^{2}}$ (orange), Ni-$d_{z^{2}}$ (green), and R-$d$ (blue). Bottom panels: Equivalent plots to those of the top row for RNiO$_2$ with R = Nd, Gd, and Ho (from left to right).
    }
    \label{fig1:Bands}
\end{figure*}

\section{\label{sec:comp}Computational Details}

The RNiO$_2$ (R = rare-earth) nickelates are the infinite layer ($n = \infty$) members of the reduced Ruddlesden-Popper series R$_{n+1}$Ni$_{n}$O$_{2n+2}$. Their crystal structure has been resolved in the \textit{P4/mmm} space group at ambient pressure \cite{crespin1983,hayward1999} (see Fig. \ref{fig:lattice}(a)). We have performed density-functional theory (DFT)-based calculations for the La-based infinite-layer nickelate LaNiO$_2$ under pressures up to 15 GPa, relevant to the experiments in Ref. \cite{wangnn_pressure_enhanced}. For all pressures, we conducted structural relaxations using the Vienna \textit{ab-initio} Simulation Package (VASP), optimizing both the lattice parameters and the internal coordinates~\cite{vasp1993,vasp1996}. The generalized gradient approximation (GGA) as implemented in the Perdew-Burke-Ernzerhof (PBE) functional was used as the exchange-correlation functional~\cite{PBE}. An energy cutoff of 500 eV and a $k$-mesh of 20$\times$20$\times$20 were adopted with a force convergence criterion of 10$^{-2}$ meV/\AA. The same structural optimization procedure was used for RNiO$_2$ (R = Nd, Pm, Eu, Gd, Ho). The lattice dynamics of RNiO$_{2}$ were further investigated using the frozen phonon method with $2\times2\times2$ supercells as implemented in \textsc{phonopy}~\cite{phonopy} interfaced with VASP.

Subsequently, for each optimized structure, we analyzed the evolution of the non-magnetic electronic structure with pressure using the all-electron, full-potential code~\textsc{wien2k}~\cite{wien2k}. In these calculations, we also employed PBE-GGA as the exchange-correlation functional. We used $R_{\mathrm{MT}}K_{\mathrm{max}}=7$ and a $k$-grid of 19$\times$19$\times$22 for the Brillouin zone sampling. Electronic structure calculations for different rare-earth ions (R = Nd, Pm, Eu, Gd, Ho) were performed using the same computational parameters within the open-core approximation for the R-$4f$ electrons. 

To gain further insight into the electronic structure and to obtain quantitative trends with pressure, we downfolded the Kohn-Sham DFT band structures onto maximally localized Wannier functions (MLWFs) to extract on-site energies and relevant hopping integrals as implemented in WANNIER90~\cite{wannier90} interfaced with WIEN2WANNIER~\cite{wien2wannier}. We chose a $d-p$ basis, where all the Ni-$3d$ and O-$2p$ orbitals were taken in the initial projections.

The magnetic tendencies of RNiO$_{2}$ under the application of hydrostatic pressure (for R = La at $P$ = 0-15 GPa) as well as for chemical pressure (R = Nd, Pm, Eu, Gd, Ho) were explored via  GGA+$U$ calculations employing the fully-localized limit (FLL) as the double counting correction \cite{PRB_dft_fll_anisimov} as implemented in VASP, with the R-$4f$ states in the pseudopotential core. Two different values for the on-site Coulomb repulsion $U$ of 2 eV and 7 eV were applied to the Ni-$3d$ states to understand the energetic dependence with $U$. The Hund's coupling $J_{\mathrm{H}}$ was fixed to the typical value of $0.7$ eV.  We considered two different antiferromagnetic (AFM) configurations that have been shown to be the lowest energy ones for RNiO$_2$ in previous DFT works \cite{Kapeghian2020, Been2021_prx}: (1) AFM-G where antiferromagnetic planes are coupled antiferromagnetically out-of-plane and (2) AFM-C where antiferromagnetic planes are coupled ferromagnetically out-of-plane.

\section{Results}

\subsection{Crystal structure and lattice dynamics}

We start by describing the changes in the crystal structure of LaNiO$_{2}$ with hydrostatic pressure. The corresponding optimized lattice parameters are shown in Fig~\ref{fig:lattice}(b). As expected, both the in-plane and out-of-plane lattice constants get reduced (by approximately 3\%) from ambient pressure to 15 GPa. We find that up to the highest pressure we studied, the space group remains as \textit{P4/mmm} (even allowing for symmetry reduction in the structural relaxations), as reflected in the calculated phonon dispersions shown in Fig. \ref{fig:lattice}(c) where no imaginary modes can be observed in the spectrum for LaNiO$_{2}$ all the way up to 15 GPa. 

To perform a meaningful comparison between the effects of hydrostatic and chemical pressure, we  make an approximate `one-to-one' mapping between the in-plane lattice constant ($a$) obtained for a certain rare-earth ion after a full structural relaxation and the corresponding pressure value applied to LaNiO$_2$. We focus on matching the in-plane lattice constant as we determined in previous work that this is the most relevant tuning knob for the electronic structure across the rare-earth series for RNiO$_2$~\cite{Kapeghian2020}. In this manner, we obtain a correspondence between NdNiO$_2$, PmNiO$_2$, EuNiO$_2$, GdNiO$_2$, and HoNiO$_{2}$ and pressures of 2, 4, 6, 8, and 15 GPa applied to LaNiO$_2$, respectively (see Fig.~\ref{fig:lattice}(b)).  The accompanying out-of-plane lattice constant for each rare-earth is slightly smaller than that obtained for the corresponding in-plane lattice-matched pressure (the optimization yields a systematic reduction of the out-of-plane lattice constant $c$ by about 10\% across the lanthanide series as can be seen in Table~\ref{tab:lattinfo} in Appendix \ref{app:crystal}). We note that it has been shown previously~\cite{PhysRevB.105.115134_dynamical_instability, PhysRevMaterials.6.044807_dynamical_instability_2} that below a critical rare-earth ionic radius, the \textit{P4/mmm} crystal structure becomes unstable in RNiO$_2$ with a lattice instability at A transforming the structure from \textit{P4/mmm} to \textit{I4/mcm} (see Fig.~\ref{fig:stability} in Appendix \ref{app:stability}). However, as mentioned above, for LaNiO$_{2}$ the \textit{P4/mmm} structure is stable for all pressures considered here. Therefore, to perform a one-to-one comparison between chemical and hydrostatic pressure in the subsequent sections, we will restrict ourselves to analyzing all RNiO$_2$ compounds in the \textit{P4/mmm} space group.

\begin{figure}
    \centering
    \includegraphics[width=0.8\columnwidth]{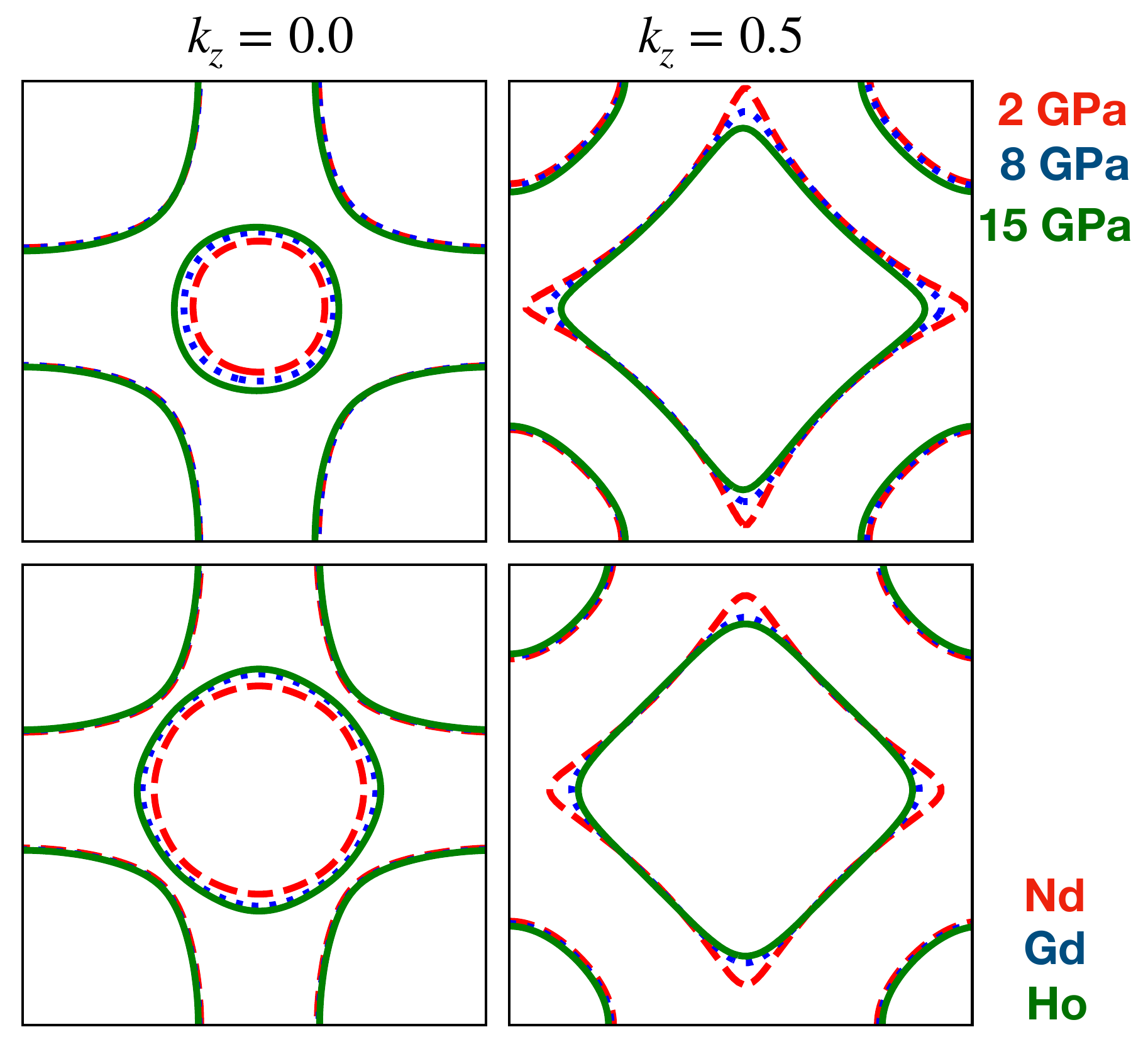}
    \caption{Top panels: Evolution of the Fermi surface of LaNiO$_2$ under hydrostatic pressure plotted in the $k_{z} = 0$ (left) and $k_{z} = 1/2$ planes (right). Bottom panels: Evolution of the Fermi surface of RNiO$_2$ (R= Nd, Gd, Ho), displaying the effects of chemical pressure  also in the $k_{z} = 0$ (left) and $k_{z} = 1/2$ planes (right). }
    \label{fig2:Fermi_Surface}
\end{figure}

\subsection{Electronic structure with pressure}
 We will start by revisiting the basic features of the non-magnetic electronic structure of LaNiO$_{2}$ at ambient pressure. As shown early on~\cite{Ni1+isnotCu2+}, three bands near the Fermi level contribute to the low-energy physics of LaNiO$_{2}$. First, a two-dimensional Ni-$3d_{x^{2}-y^{2}}$-derived band crosses the Fermi level, reminiscent of the cuprates. But, as mentioned above, the infinite-layer nickelates host additional La-$5d$ `spectator' bands which self-dope the Ni-$3d_{x^{2}-y^{2}}$ orbital. These La-$5d$ bands generate two electron pockets: one at $\Gamma$ with mostly La-$d_{z^{2}}$ character and one at A with mostly La-$d_{xy}$ character. Importantly, the $5d$ bands have a three-dimensional dispersion giving rise to strong out-of-plane couplings~\cite{jung2022antiferromagnetic,labollita2023conductivity}. The self-doping effect (that corresponds to around 5\% holes in the Ni-$d_{x^{2}-y^{2}}$ orbital) causes a shift away from nominally half-filling. Therefore, the infinite-layer nickelates can be thought of as equivalent to underdoped cuprates even at stoichiometry (i.e. nominal $d^9$ filling).  Further, when compared to cuprates, the O-$2p$ bands in infinite-layer nickelates are much lower in energy relative to the Ni-$3d$ bands, giving rise to a $\sim 4.3$ eV charge-transfer energy, which is much larger than typical  cuprate values $\sim1-2$ eV \cite{Weber_2012}. In this context, an important question in infinite-layer nickelates has been their placement on the charge transfer-Mott continuum defined by Zaanen, Sawatzky and Allen \cite{ZSA_phase_diagram}. A recent RIXS study \cite{dean2022} suggests that the reduced nickelates are intermediate between the charge transfer and Mott limits.

We now turn to the electronic structure of LaNiO$_{2}$ under the influence of hydrostatic and chemical pressure. Fig.~\ref{fig1:Bands} shows the band structure evolution -- with the relevant orbital characters highlighted -- as well as the corresponding atom and orbital-projected density of states (DOS) up to 15 GPa for LaNiO$_2$ as well as for RNiO$_2$ with R = Nd (matching 2 GPa), R = Pm (matching 8 GPa) and R = Ho (matching 15 GPa) (further pressures and R ions are shown in Fig. \ref{fig:P_0_4_6_bands} in Appendix \ref{app:dft}). We focus first on the evolution of the Ni-$d_{x^{2}-y^{2}}$ band that crosses the Fermi level in LaNiO$_{2}$  with hydrostatic pressure: its bandwidth increases monotonically by about 5\% with pressures up to 15 GPa. This is expected considering the trend in lattice constants shown in the previous section: a reduction of the in-plane constants leads to increased orbital overlap and, as a consequence, in the hybridization and bandwidth. A similar increase in bandwidth is observed when reducing the size of the rare-earth ion from La to Ho. 

The electron pockets associated with the rare-earth $5d$ bands that cross the Fermi level can be seen to increase in size with pressure in Fig.~\ref{fig1:Bands}. To further elucidate this trend, we also show the corresponding Fermi surfaces as a function of pressure in the $k_{z}=0$ and $k_{z}=1/2$ planes in Fig.~\ref{fig2:Fermi_Surface}. While the size of the large hole-like Ni-$d_{x^{2}-y^{2}}$ pocket does not change significantly with pressure or rare-earth size, the size of the electron pocket with R-$d_{z^2}$ character around the $\Gamma$-point can been seen to gradually increase in $k_{z}=0$ with pressure (the R-$d_{xy}$ pocket at the zone corners (A) in $k_{z}=1/2$ remains essentially unchanged with pressure instead). While the effects on the R-$5d$ pockets with hydrostatic and chemical pressure are similar, the latter systematically gives rise to a larger electron pocket at $\Gamma$. In any case, for both hydrostatic and chemical pressure, the largest effect on the fermiology seems to be an increase in the amount of self-doping of the Ni-$d_{x^{2}-y^{2}}$ band as the R-$5d$ pockets become enlarged.

\begin{figure}
    \centering
    \includegraphics[width = 0.95\columnwidth]{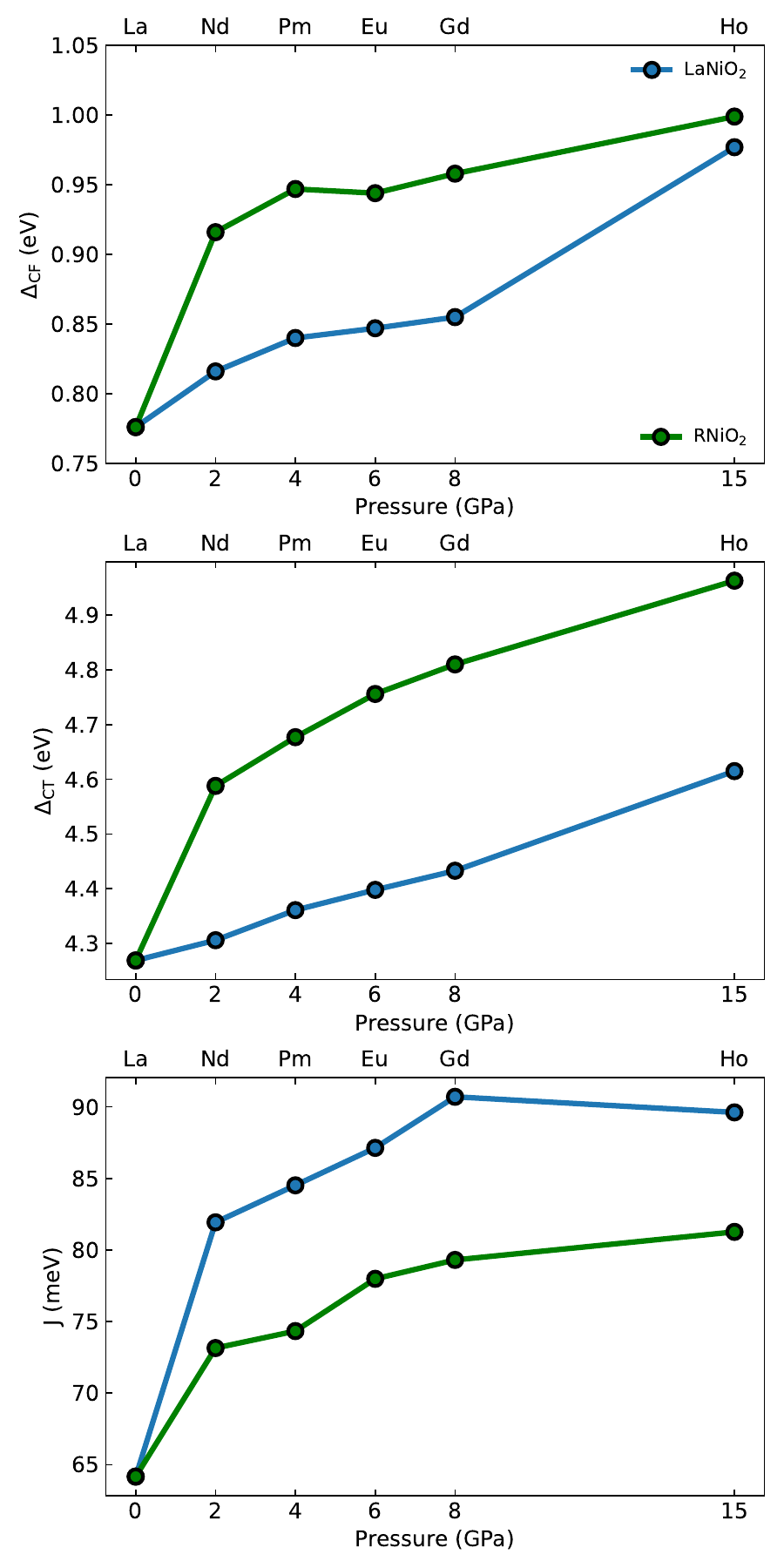}
    \caption{Quantitative trends in the electronic structure of RNiO$_{2}$ with hydrostatic and chemical pressure obtained from MLWFs: (from top to bottom) crystal-field splitting ($\Delta_{\mathrm{CF}}$) between the Ni-$e_{g}$ orbitals, charge-transfer energy ($\Delta_{\mathrm{CT}}$) and estimated superexchange ($J$). The blue curve shows the  evolution of each of these parameters in LaNiO$_{2}$ for pressures up to 15 GPa, the green curve shows the evolution of these parameters for RNiO$_{2}$ (R= Nd, Pm, Eu, Gd, and Ho).}
    \label{fig:superexchange}
\end{figure}

Moving to the changes in $p-d$ hybridization, the complex of O-$2p$ bands can be observed to shift down in energy with increasing pressure (both hydrostatic and chemical) while the Ni-$3d$ states do not significantly move, consequently decreasing the degree of $p-d$ hybridization  (see Fig. \ref{fig1:Bands}). This is an unexpected effect that has been described before when analyzing the effects of chemical pressure in the infinite-layer nickelates~\cite{Kapeghian2020, Been2021_prx} (given that the lattice constants contract while the bandwidth increases one would in principle expect that the degree of covalency associated with the holes in the Ni-O planes should also increase). With these considerations, to obtain an estimate of the change in the degree of $p-d$ hybridization with pressure, we calculate the charge-transfer energy evolution $\Delta_{\mathrm{CT}} = \varepsilon_{d}-\varepsilon_{p}$ (referring to $d_{x^2-y^2}$ and $p\sigma$) from the on-site energies of MLWFs  (see relevant Wannier fits in Figs. \ref{Appendix:wannier_bands} and \ref{Fig:Rare_Earth_Wannier_Bands} and on-site energies and hopping integrals in Table~\ref{tab:wannier} in Appendix \ref{app:wannier}). The derived values of $\Delta_{\mathrm{CT}}$ with pressure are shown in Fig. \ref{fig:superexchange} where $\Delta_{\mathrm{CT}}$ can be seen to increase from 4.27 eV at ambient pressure up to 4.62 eV at 15 GPa in LaNiO$_2$. A slightly larger increase is obtained when changing the rare-earth size, with $\Delta_{\mathrm{CT}}$ increasing from 4.59 eV to 4.96 eV when going from NdNiO$_{2}$ to HoNiO$_{2}$.

Using the on-site energies from MLWFs (see Table \ref{tab:wannier} in Appendix \ref{app:wannier}), we also analyze the evolution of the Ni-$e_{g}$ crystal-field splitting defined as $\Delta_{\mathrm{CF}} = \varepsilon_{d_{x^{2}-y^{2}}}-\varepsilon_{d_{z^{2}}}$ that can be seen to systematically increase with both hydrostatic and chemical pressure from $\sim$ 0.78 eV at ambient pressure for LaNiO$_2$ to $\sim$ 1 eV for 15 GPa and HoNiO$_2$  (see Figure~\ref{fig:superexchange}). As mentioned above, a larger value of $\Delta_{\mathrm{CF}}$ in the cuprates is correlated with a higher T$_c$ due to the reduced mixing of these
orbitals~\cite{Sakakibara2014, sakakibara_splitting}.

\subsection{Magnetic tendencies with pressure}
\begin{figure}
    \centering
    \includegraphics[width  = \columnwidth]{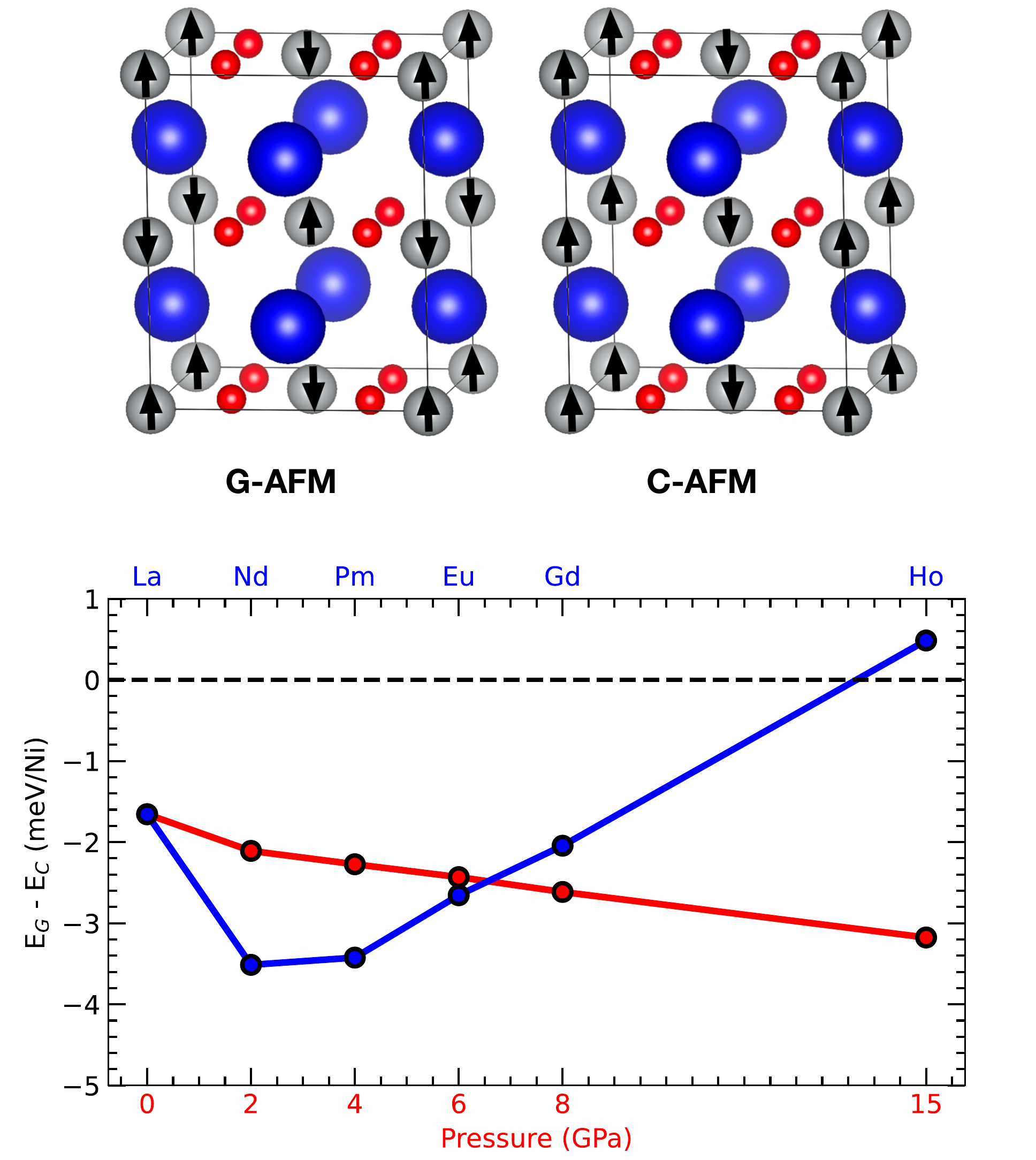}
    \caption{Magnetic tendencies of RNiO$_{2}$ with hydrostatic and chemical pressure. Top panel: AFM-G (left) and AFM-C (right) spin configurations. Bottom panel: Energy difference between the G-type and C-type antiferromagnetic states with hydrostatic and chemical pressure within GGA-PBE+$U$ ($U$= 2 eV) for RNiO$_2$. The curve in red shows the energy difference for LaNiO$_{2}$ for pressures up to 15 GPa, the blue curve shows the same energy difference for RNiO$_{2}$ (R= Nd, Pm, Eu, Gd, and Ho). The G-type AFM state is the ground state throughout except for Ho.}
    \label{fig:magnetic_energy_diff}
\end{figure}

In cuprates, the ground state of the undoped compounds (at nominal $d^9$ filling) is an antiferromagnetic charge-transfer insulator \cite{Keimer2015}. In infinite-layer nickelates the situation is slightly different: as mentioned above, there is evidence of short-range antiferromagnetic fluctuations from RIXS~\cite{Lu2021} even though long-range magnetic order has not been confirmed, likely due to the interference of the R-$5d$ bands. Regardless, within DFT (and DFT+$U$) calculations, an antiferromagnetic ground state is obtained at ambient pressure~\cite{Ni1+isnotCu2+, botana2020, Been2021_prx}. Hence, to better understand the magnetic tendencies with pressure, we perform spin-polarized calculations and compare the energies of the two most stable magnetic states in LaNiO$_{2}$ at ambient pressure (G-type AFM and C-type AFM). Fig. \ref{fig:magnetic_energy_diff} shows the evolution of the energy difference between these two magnetic states with pressure (hydrostatic and chemical) at $U$= 2 eV (results for $U$= 7 eV are shown in Fig. \ref{fig:G_C_AFM_U_7} in Appendix \ref{app:spin}). For LaNiO$_2$ the G-type AFM state is the ground state up to 15 GPa (and it is further stabilized as pressure is increased). For RNiO$_{2}$, as we move to the right in the lanthanide series, the G-AFM state continues being the ground state up to Gd, becoming less stable as chemical pressure is increased. Eventually, a crossover occurs, so that for Ho the magnetic ground state is C-type AFM. At $U$= 7 eV, the G-type AFM state becomes even further stabilized, as shown in Fig. \ref{fig:G_C_AFM_U_7} in Appendix \ref{app:spin}. 

As mentioned above, in cuprates there is a general consensus that magnetism is a key aspect underlying their physics, with the energy scale of T$_c$ being set by the large values of the superexchange ($J$). We will also use the scale of $J$ here as a predictor of the superconducting tendencies in the infinite-layer nickelates with pressure. One can in principle extract estimates for the exchange constants from DFT-energy maps to a Heisenberg model. However, for RNiO$_2$ the moments obtained in some of the magnetic solutions disproportionate, implying that such a mapping is not appropriate. Hence, to estimate the evolution of the magnetic superexchange with pressure, we resort to using both the Mott- and charge-transfer limits in the Zaanen-Sawatzky-Allen phase diagram \cite{ZSA_phase_diagram} (given that in these materials $U \sim \Delta_{\mathrm{CT}}$),
\begin{equation*} 
    J =  \frac{2t_{pd}^{4}}{\Delta^{2}_{\mathrm{CT}}} \times ( \frac{1}{U_{dd}}+ \frac{1}{\Delta_{\mathrm{CT}} + \frac{1}{2}U_{pp}})
\end{equation*}
Using our estimates for $\Delta_{\mathrm{CT}}$ and $t_{pd}$ derived from the wannierizations, together with values of U$_{pp} = 7.35$ eV and U$_{dd} = 6.00$ eV, characteristic of the cuprates \cite{Cuprates_coulomb_U_J}, we obtain a $J$ value for LaNiO$_2$ at ambient pressure $\sim$ 65 meV (very similar to the experimental value obtained from fittings of the spin wave dispersion derived from RIXS ~\cite{Lu2021}). Importantly, we find that $J$ increases to a cuprate-like value of $\sim 90$ meV at a modest pressure of 8 GPa, as shown in Fig.~\ref{fig:superexchange}. Even though we restricted ourselves here to pressures relevant to the experiments in Ref. \cite{wangnn_pressure_enhanced} ($\sim$ 12 GPa), in principle an even larger $J$ could be achieved at higher pressures (see Fig. \ref{fig:superexchange_full_press} in Appendix \ref{app:additional_super}). This large increase in the superexchange value (together with the main electronic structure trends described above) agree well with previous theoretical calculations~\cite{dicataldo2023unconventional}, that have predicted an increase in the superconducting T$_{c}$ of the hole-doped Pr infinite-layer nickelate with pressure.

\section{\label{sec:disc}Summary and Conclusions}
We have investigated the evolution of the electronic structure and magnetic trends in infinite layer nickelates ($R$NiO$_2$) with pressure using first-principles calculations. Our findings suggest that there is a one-to-one correspondence between hydrostatic and chemical pressure in these materials. Overall, our results show that pressure (both hydrostatic and chemical)  tends to (i) increase the R-$5d$ self-doping effect, (ii) increase the Ni-$d_{x^{2}-y^{2}}$ bandwidth, (iii) increase the $e_g$ splittings, (iv) decrease the degree of $p-d$ hybridization, and (v) increase the superexchange ($J$). Using the energy scale of $J$ as a predictor of superconducting tendencies, we anticipate that hydrostatic pressure and rare-earth substitution can indeed be a feasible means to further increase the T$_c$ in this family of materials.

\begin{acknowledgements}
We acknowledge NSF Grant No. DMR-2045826 and the ASU Research Computing Center for HPC resources. 
\end{acknowledgements}


\onecolumngrid

\appendix

\section{\label{app:crystal}Additional structural data for RNiO$_{2}$}
Table~\ref{tab:lattinfo} summarizes the DFT-optimized out-of-plane ($c$) lattice constants for pressurized LaNiO$_{2}$ and the corresponding in-plane lattice-matched RNiO$_{2}$. 
 
\begin{table}[h]
\begin{tabular*}{0.6\columnwidth}{c@{\extracolsep{\fill}}cccccc}
\hline\hline
             & 0 GPa  & 2 GPa  & 4 GPa  & 6 GPa & 8 GPa & 15 GPa \\
             \hline
    $c (\text{\AA}) $ & 3.371  & 3.369 & 3.353 & 3.334 & 3.316 & 3.262 \\
    \hline
        & La     & Nd    & Pm     & Eu    & Gd     & Ho\\
        \hline
    $c (\text{\AA})$  & 3.371  & 3.302  & 3.263 & 3.206 & 3.177  & 3.110\\
   \hline\hline
\end{tabular*}
\caption{Optimized $c$ lattice constants under applied pressure for LaNiO$_2$ and for RNiO$_2$ with R = Nd, Pm, Eu, Gd, and Ho. All the lattice constants are provided in \AA.}
\label{tab:lattinfo}
\end{table}

\section{\label{app:stability}Dynamical instability of RNiO$_{2}$}

\begin{figure*}
    \centering
    \includegraphics[width=0.7\columnwidth]{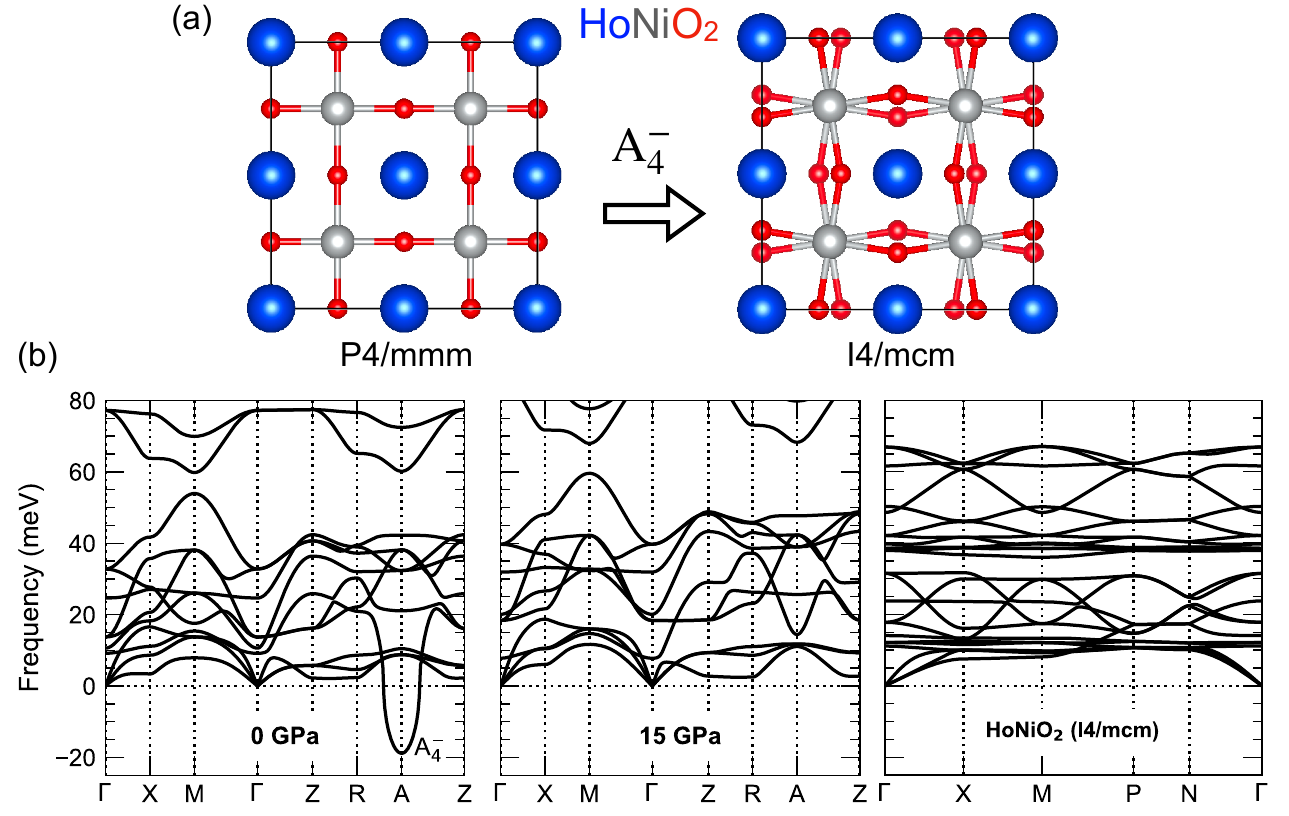}
    \caption{Dynamical instability of RNiO$_{2}$ infinite-layer nickelates. R = Ho is taken as an example. (a) RNiO$_{2}$ crystal structure in the high-symmetry ($P4/mmm$) phase (left) and low-symmetry, distorted phase ($I4/mcm$) (right). (b) Phonon dispersions for HoNiO$_{2}$ at ambient pressure (left) and 15 GPa (center) for the $P4/mmm$ crystal structure. The soft-phonon mode at the A ($\mathbf{q}=(1/2,1/2,1/2)$) point corresponds to an A$_{4}^{-}$ normal mode. Phonon dispersion for HoNiO$_{2}$ at ambient pressure in the $I4/mcm$ space group with no imaginary phonon modes (right).}
    \label{fig:stability}
\end{figure*}

As mentioned in the main text, it has been shown previously~\cite{PhysRevB.105.115134_dynamical_instability, PhysRevMaterials.6.044807_dynamical_instability_2} that above a critical rare-earth ionic radius, the \textit{P4/mmm} crystal structure becomes unstable in RNiO$_2$ with a lattice instability at A transforming the space group from \textit{P4/mmm} to \textit{I4/mcm} (see Fig.~\ref{fig:stability}(a)). However, as mentioned above, for LaNiO$_{2}$ the \textit{P4/mmm} structure is stable for all pressures considered here (see Fig.~\ref{fig:lattice}(b)). For smaller rare-earth ions, we take Ho as an example and reproduce the results of previous work~\cite{PhysRevB.105.115134_dynamical_instability, PhysRevMaterials.6.044807_dynamical_instability_2}, where a lattice instability appears at A (see Fig.~\ref{fig:stability}(b)). This unstable mode corresponds to the A$_{4}^{-}$ irreducible representation of the \textit{P4/mmm} space group and leads to a crystal structure with rotated oxygens (\textit{I4/mcm} symmetry). The distorted \textit{I4/mcm} structure is dynamically stable as shown in Fig.~\ref{fig:stability}(b) evidenced by the lack of imaginary phonon modes. However, if one applies hydrostatic pressure to HoNiO$_{2}$ we note that this lattice instability can be quenched (see Fig.~\ref{fig:stability}(b)). 

\section{\label{app:dft}Additional DFT data for RNiO$_{2}$ with hydrostatic and chemical pressure in the nonmagnetic state}
\begin{figure*}
    \centering
    \includegraphics[width = \columnwidth]{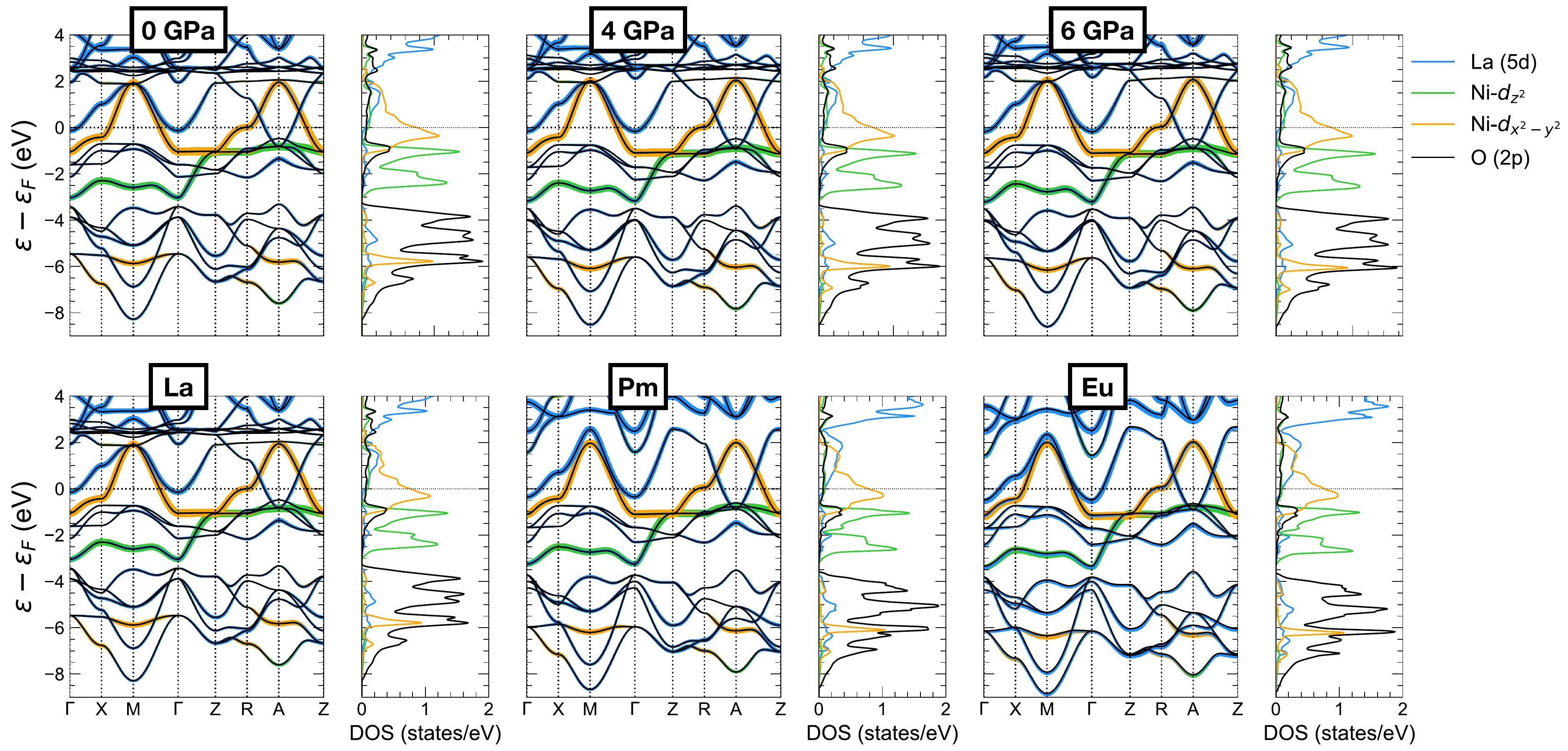}
    \caption{Comparison between the non-magnetic electronic structure of RNiO$_2$ with hydrostatic (top) and chemical (bottom) pressure. Top panels: Band structure along high-symmetry lines and atom-, orbital-resolved density of states (DOS) for LaNiO$_2$ at ambient pressure, 4 GPa, and 6 GPa (from left to right). The orbital character of the bands is denoted for Ni-$d_{x^{2}-y^{2}}$ (orange), Ni-$d_{z^{2}}$ (green), and R-$d$ (blue). Bottom panels: Equivalent plots as those of the top row for RNiO$_2$ with R = La, Pm, and Eu (from left to right).
    }
    \label{fig:P_0_4_6_bands}
\end{figure*}

Figure~\ref{fig:P_0_4_6_bands} summarizes the non-magnetic electronic structure for LaNiO$_{2}$ at additional pressures, and the corresponding lattice-matched rare-earth cation.
The same trends as those described in the main text can be observed. 

\section{\label{app:wannier}Wannierization of the DFT bands}
\begin{figure*}
    \centering
    \includegraphics[width = 0.85\columnwidth]{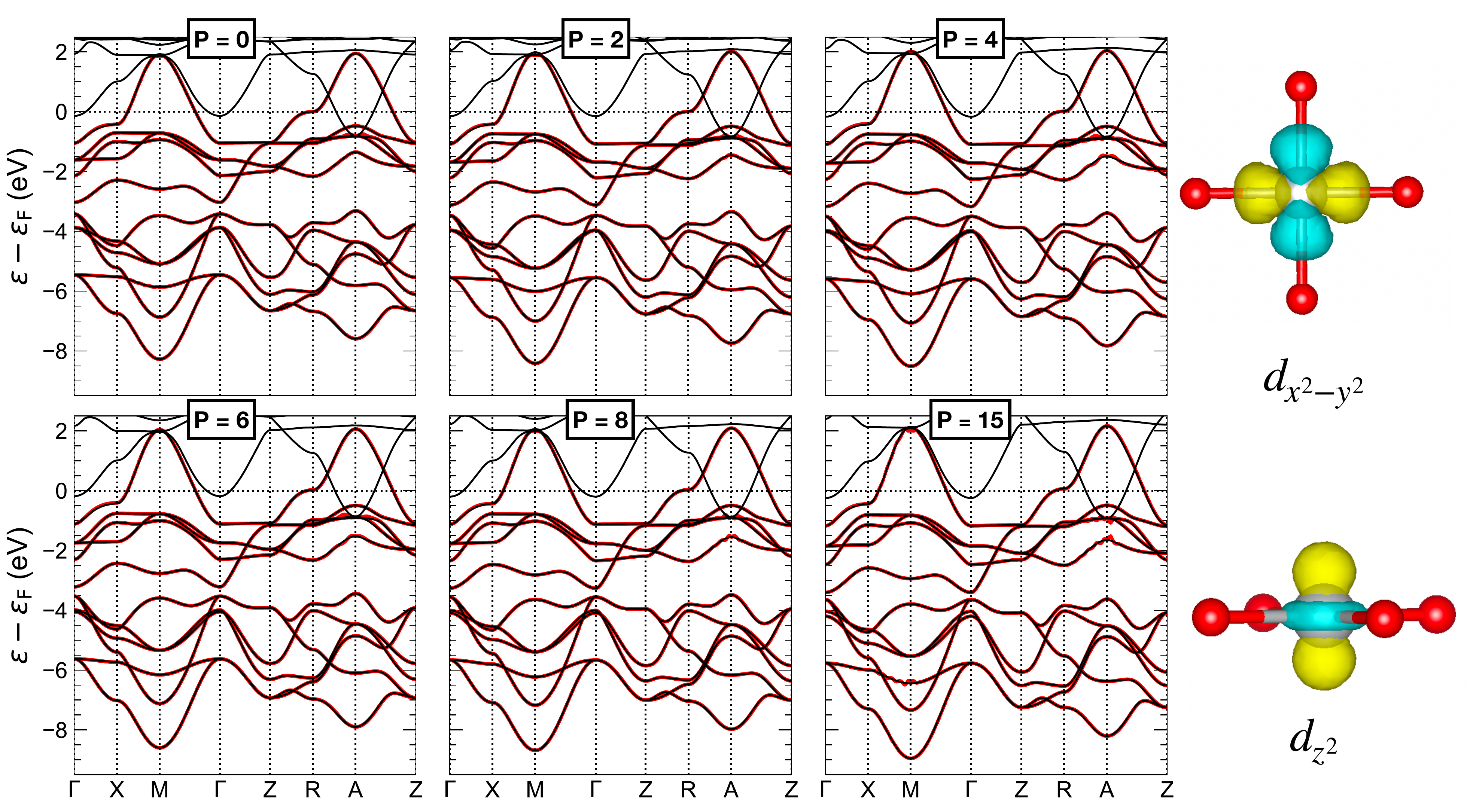}
    \caption{Left panel: Comparison between the DFT bands (black) and the bands derived from our $d-p$ MLWFs (red) for LaNiO$_2$ for pressures up to 15 GPa. Right panel: Real-space visualization of the localized, atomic-like MLWFs corresponding to the Ni-$e_{g}$ orbitals at ambient pressure.}
    \label{Appendix:wannier_bands}
\end{figure*}

\begin{figure*}[h]
    \centering
    \includegraphics[width = 1.0\textwidth]{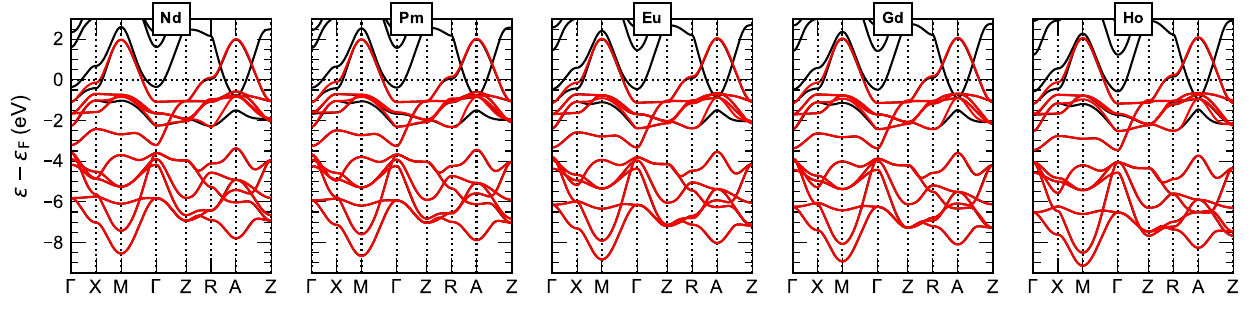}
    \caption{Comparison between the DFT bands (black) and the bands derived from our $d-p$ MLWFs (red) for RNiO$_2$ R= Nd, Pm, Eu, Gd, and Ho.}
    \label{Fig:Rare_Earth_Wannier_Bands}
\end{figure*}

To obtain a more quantitative analysis of the DFT electronic structure, we downfold the DFT Kohn-Sham bands onto a $d-p$ basis of MLWFs. While this procedure does not result in a unique basis, we find that our downfolding provides a faithful representation of the DFT bands. We compare our DFT bands to the MLWFs-derived bands in  Fig. ~\ref{Appendix:wannier_bands} (for different pressures applied to LaNiO$_2$) and Fig. \ref{Fig:Rare_Earth_Wannier_Bands} (for different rare-earths) where excellent agreement can be observed. Our MLWFs are well-localized and atomic-like as indicated by the real space visualization of the MLWFs corresponding to the Ni-$d_{x^{2}-y^{2}}$ and Ni-$d_{z^{2}}$ orbitals (see Fig. ~\ref{Appendix:wannier_bands}).

On-site energies and relevant hopping integrals obtained from the Wannier Hamiltonians for different hydrostatic pressures and rare-earth cations are summarized in Table~\ref{tab:wannier} (all quantities in units of eV). Note that O-$p_{\sigma(\pi)}$ corresponds to the in-plane bonding (anti-bonding) with the Ni-$d_{x^{2}-y^{2}}$ orbital. The in-plane hopping is estimated from $t_{pd}$ which is the hopping integral from the Ni-$d_{x^{2}-y^{2}}$ to the bonding O-$p_{\sigma}$ orbital. The effective $t_{dd}$ hopping can be obtained using $t_{dd} = t_{pd}^{2}/\Delta_{\mathrm{CT}}$. From this data, we derive the charge-transfer energy, Ni-$e_{g}$ splitting, as well as estimate the superexchange $J$, as discussed in the main text.

\begin{table*}
    \centering
    \begin{tabular*}{0.9\columnwidth}{l@{\extracolsep{\fill}}cccccc}
    \hline\hline
                                  & 0 GPa & 2 GPa & 4 GPa & 6 GPa & 8 GPa & 15 GPa \\
    \hline
    Wannier on-site energies (eV)  &       &       &       &       &       &  \\
    \hline
    Ni-$d_{x^{2}-y^{2}}$ & $-1.094$ & $-1.138$  & $-1.147$ & $-1.163$ & $-1.181$  & $-1.172$  \\
    Ni-$d_{z^{2}}$       & $-1.870$ & $-1.954$  & $-1.987$ & $-2.010$ & $-2.036$  & $-2.150$ \\
    Ni-$d_{xy}$          & $-1.748$ & $-1.803$  & $-1.840$ & $-1.874$ & $-1.911$  & $-1.982$ \\
    Ni-$d_{yz/xz}$       & $-1.632$ & $-1.701$  & $-1.733$ & $-1.760$ & $-1.787$  & $-1.876$ \\
    O-$p_{\pi}$          & $-4.485$ & $-4.547$  & $-4.589$ & $-4.623$ & $-4.659$  & $-4.782$ \\
    O-$p_{\sigma}$       & $-5.363$ & $-5.444$  & $-5.508$ & $-5.561$ & $-5.614$  & $-5.789$ \\
    O-$p_{z}$            & $-4.485$ & $-4.567$  & $-4.606$ & $-4.639$ & $-4.674$  & $-4.764$ \\
    \hline
    Wannier hoppings (eV) &         &            &         &          &           &          \\
    \hline
    Ni-$d_{x^{2}-y^{2}}$-O-$p_{\sigma}$ &  $-1.189$ & $-1.270$   & $-1.289$ & $-1.305$ & $-1.324$  & $-1.347$ \\
    Ni-$d_{z^{2}}$ - Ni-$d_{z^{2}}$ (001) & $-0.365$ & $-0.360$  & $-0.367$  & $-0.377$ & $-0.386$ & $-0.389$ \\
    \hline\hline
                           &   & NdNiO2 &PmNiO2 &  EuNiO2 & GdNiO2 & HoNiO2  \\
    \hline
    Wannier on-site energies (eV)       & &  & & &  &  \\
    \hline
    Ni-$d_{x^{2}-y^{2}}$    & & $-0.998$ & $-0.992$ & $-1.037$ & $-1.049$ & $-1.058$\\
    Ni-$d_{z^{2}}$          & & $-1.914$ & $-1.939$ & $-1.981$ & $-2.007$ & $-2.057$\\
    Ni-$d_{xy}$             & & $-1.785$ & $-1.813$ & $-1.882$ & $-1.914$ & $-1.981$ \\
    Ni-$d_{yz/xz}$           & & $-1.636$ & $-1.660$ & $-1.735$ & $-1.767$ & $-1.820$   \\
    O-$p_{\pi}$            & & $-4.847$ & $-4.964$ & $-5.170$ & $-5.271$ & $-5.540$  \\
    O-$p_{\sigma}$         & & $-5.586$ & $-5.669$ & $-5.793$ & $-5.859$ & $-6.021$  \\
    O-$p_{z}$              & & $-4.810$ & $-4.901$ & $-5.064$ & $-5.142$ & $-5.342$   \\
    \hline
    Wannier hoppings (eV) & & & & & &  \\
    \hline
    Ni-$d_{x^{2}-y^{2}}$-O$_{p_{\sigma}}$ & & $-1.279$ & $-1.298$ & $-1.326$ & $-1.340$ & $-1.372$  \\
    Ni-$d_{z^{2}}$-Ni-$d_{z^{2}}$ (001)   & & $-0.420$ & $-0.436$ & $-0.459$ & $-0.469$ & $-0.500$   \\
    \hline\hline
    \end{tabular*}
    \caption{Summary of the calculated on-site energies and hopping integrals from Wannier functions for pressurized LaNiO$_2$ and RNiO$_2$ (R= Nd, Pm, Eu, Gd, Ho). O-$p_{\pi(\sigma)}$ denotes the anti-bonding (bonding) O-$p$ orbital in the NiO$_2$ plane. All quantities are in units of eV.}
    \label{tab:wannier}
\end{table*}

\begin{figure*}
    \centering
   \includegraphics[width = 0.6\columnwidth]{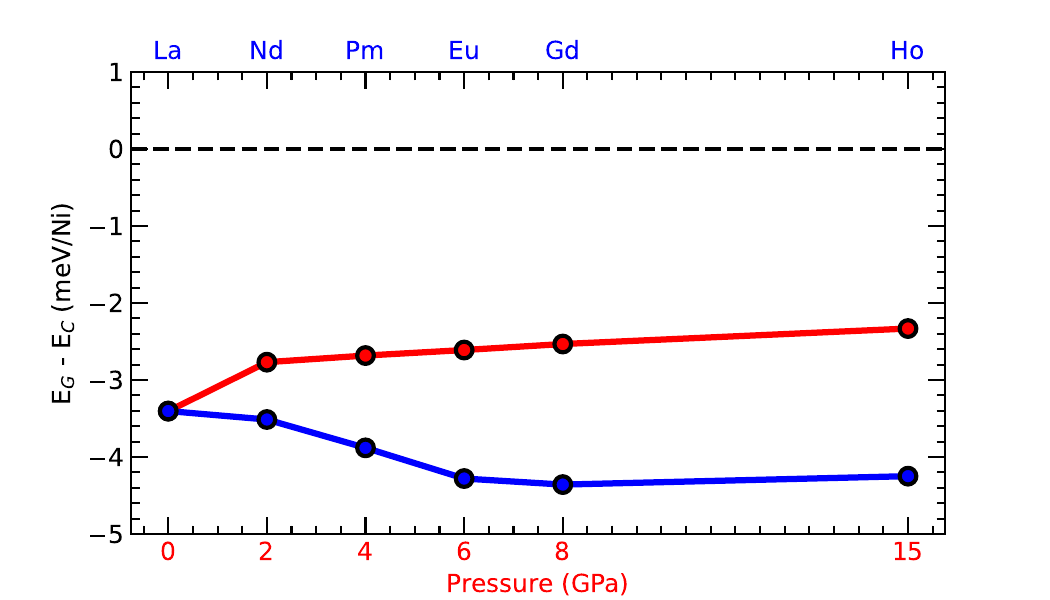}
    \caption{Evolution of the energy difference between the G-type and C-type AFM states in RNiO$_2$ with hydrostatic (red curve) and chemical (blue curve) pressure at $U$ = 7 eV.}
    \label{fig:G_C_AFM_U_7}
\end{figure*}

\section{\label{app:spin}Additional DFT data for spin-polarized calculations in RNiO$_{2}$ with hydrostatic and chemical pressure}
Fig. \ref{fig:G_C_AFM_U_7} shows the evolution with pressure of the energy difference between a C-type and a G-type AFM state at $U$= 7 eV ($J = 0.7$ eV). The G-type AFM state is more stable for all the pressures and rare-earth ions studied here. The energetics between AFM-C and AFM-G solutions vary throughout the literature (albeit the energy scales are rather small) indicating competition between aligning or anti-aligning the Ni moments out-of-plane~\cite{Kapeghian2020, Ni1+isnotCu2+,Liu2020}.

\section{\label{app:additional_super} Superexchange strength for a broader pressure range}

\begin{figure*}[h!]
    \centering
    \includegraphics[width = 0.5\columnwidth]{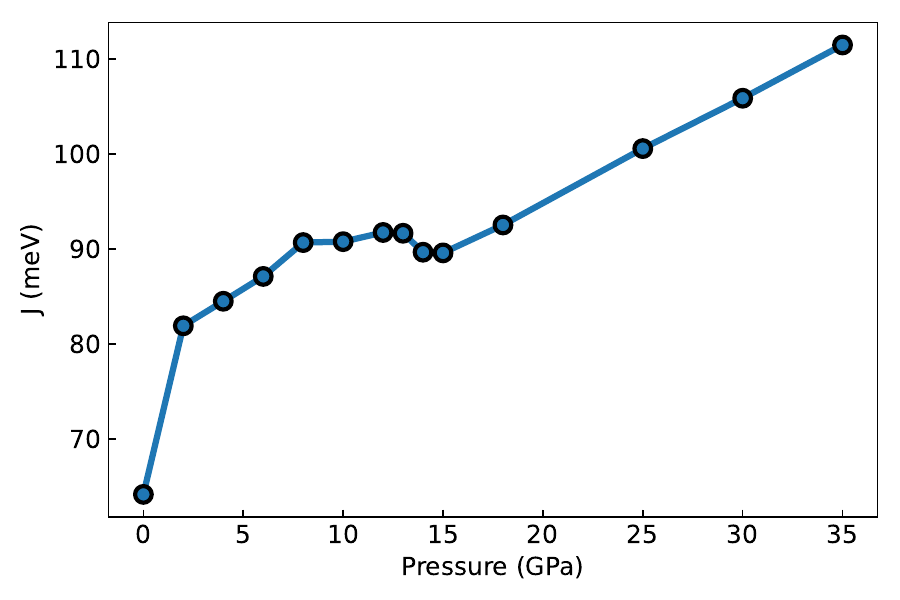}
    \caption{Evolution of the superexchange coupling ($J$) of LaNiO$_{2}$ under hydrostatic pressures up to 35 GPa.}
    \label{fig:superexchange_full_press}
\end{figure*}

Fig. \ref{fig:superexchange_full_press} shows the evolution of the superexchange ($J$) of LaNiO$_{2}$ under hydrostatic pressure (up to 35 GPa). Large cuprate-like values ($\sim$ 110 meV) can be achieved at the largest applied pressure of 35 GPa.

\twocolumngrid

\clearpage\newpage


%

\end{document}